\documentclass[journal,comsoc]{IEEEtran}
\usepackage[justification=centering]{caption}

\usepackage{subfigure}

\usepackage{graphicx}
\usepackage[export]{adjustbox}

\usepackage{float}
\ifCLASSOPTIONcompsoc
 \usepackage[nocompress]{cite}
\else
 \usepackage{cite}
\fi
\usepackage{fixltx2e}

\usepackage{graphicx}

\usepackage{dblfloatfix}
\usepackage{rotating}
\usepackage{booktabs}
\usepackage{multirow}
\usepackage[table]{xcolor}
\usepackage{longtable}
\usepackage{lscape}
\usepackage{amsmath, amsthm, amssymb}
\usepackage{url}
\usepackage{lipsum}
\usepackage[T1]{fontenc} 
\usepackage{enumitem}
\usepackage[utf8]{inputenc}
\usepackage{balance}

\usepackage{url}

\usepackage{algpseudocode}
\usepackage[linesnumbered,ruled,vlined]{algorithm2e}
\usepackage{subfigure}
\usepackage{booktabs} 
\usepackage{float}
\usepackage{color}

\usepackage{authblk}

\vspace{10pt}
\author[1,2]{Guilherme B. Araujo}
\author[1]{Maycon L. M. Peixoto}
\author[1]{Leobino N. Sampaio}
\author[ ]{\\ \vspace{10pt} \texttt{guilherme.araujo@ufrb.edu.br,} \texttt{\{maycon.leone, leobino\}@ufba.br}}

\affil[1]{Federal University of Bahia (UFBA)}
\affil[2]{Federal University of Rec\^oncavo da Bahia (UFRB)}

\title{NDN4IVC: A Framework for Simulating and Testing of Applications in Vehicular\\ Named Data Networking}

\begin{document}


\IEEEcompsoctitleabstractindextext{%
\begin{abstract}
This paper presents a customized framework (NDN4IVC) for simulating and testing intelligent transportation systems and applications in vehicular named-data networking (V-NDN). The project uses two popular simulators in the literature for VANET simulation, a network simulator based on discrete events (Ns-3), with ndnSIM module installed, and Sumo, a simulator for urban mobility. NDN4IVC allows bidirectional communication between Sumo and Ns-3 and integrates the NDN stack and the NFD (NDN Forwarding Daemon) code. The project also brings together a comprehensive set of codes, models, functionalities, and technologies to improve proposals and protocols in V-NDN.
\end{abstract}



}

\maketitle


\IEEEdisplaynotcompsoctitleabstractindextext

%


\section{Introduction} \label{sec:introduction} 

Realistic simulation of Inter-Vehicle Communication (IVC) is one of the main challenges in the vehicular networks (Vehicular Ad Hoc Networks -- VANETs) research domain \cite{Martinez2009}. Some network architectures can be used and adapted for this purpose \cite{10.1007/978-3-642-29222-4_38}. 

Named data networking (NDN) is an information-centric network architecture designed to offer cache in the network layer, native multicast and multihoming support, data security, mobility support, traffic control, among other benefits. Each NDN node maintains three data structures to offer these services: PIT (Pending Interest Table), FIB (Forwarding Information Base), and CS (Content Store). The CS is a temporary cache of Data packets that the node has received, which can potentially be used to satisfy future Interests. The PIT stores all Interests that have been forwarded but not satisfied yet. If a received Interest does not match either the CS or the PIT, it will be forwarded toward the data producer(s) according to the FIB. When a Data packet returns, the node finds the matching PIT entry, forwards the Data to all downstream interfaces listed in the PIT entry, and then removes that entry and caches the Data. When a Data packet does not match any PIT entry, it will be considered unsolicited and dropped \cite{10.1145/2656877.2656887}, \cite{6849267}.

In VANETs, due to the specificity of the environment with high node mobility, constant topology changes, and frequent disconnections, the NDN structures and services can help efficient data delivery for distributed applications. Moreover, in vehicular named-data networking (V-NDN), vehicles can play the following roles: (i) data consumer, (ii) content producer, (iii) forwarder, when connected to infrastructure or other vehicles, and (iv) ``data mule'', when transporting data even without network connectivity, spreading content between different areas \cite{6849267}. Moreover, unlike other mobile devices, vehicles move over known road infrastructure and don't care about storage capacity.

There are works in the literature that use tools and simulators to evaluate proposals in V-NDN that are not suitable and customized environments for VANET simulation. In some cases, it is not possible to obtain information about the mobility of vehicles on the road infrastructure. Veins\footnote{http://veins.car2x.org}, a VANET simulation framework, uses the Sumo\footnote{https://www.eclipse.org/sumo} as an urban mobility simulator and OMNeT++\footnote{https://omnetpp.org/} as a network simulator. The framework implements the WAVE (Wireless Access to Vehicular Environment) architecture \cite{10.1007/978-3-642-29222-4_38} for inter-vehicle communications. However, performing simulations in Veins of proposals in V-NDN is not a trivial task. It requires the user to implement the entire packet structure, storage, and forwarding mechanism of NDN architecture. On the other hand, ndnSIM\footnote{https://ndnsim.net} \cite{10.1145/3138808.3138812}, a module for the Ns-3 network simulator, brings the implementation of the entire NDN stack to the simulation environment. Thus, the module allows working with packages in name format through the \texttt{ndn-cxx}\footnote{https://named-data.net/doc/ndn-cxx} library (NDN C++ library with eXperimental eXtensions). In addition, all management and processing for forwarding packets on the network are done through the source code of the NFD (Named data networking Forwarding Daemon) \cite{ndnGitReposiories}. However, the ndnSIM cannot be classified as a VANET simulator, as discussed in \cite {Martinez2009}. That's because the environment has limited information about the road infrastructure and node's mobility on their respective lanes and routes.

The lack of environments designed for evaluating V-NDN applications can make it impossible to reproduce more realistic scenarios or generate inconsistent research results \cite{8624354}. Furthermore, the limitation of existing environments leads to the proposition of solutions for VANETs that do not exploit all the properties of the NDN architecture, which could benefit the development of applications and solve problems intrinsic to the mobility environment of VANETs. Therefore, given the lack of an adequate environment for V-NDN simulation \cite{8624354}, the main objective of this work is: present the framework NDN4IVC (NDN for Inter-Vehicular Communication) to permit application experiments in VANETs with native support of the NDN architecture.

This article is organized as follows: Section \ref{sec:proposal} introduces the framework. Then, the experiments and results are discussed in Section \ref{sec:avaliacao}. Section \ref{sec:roteiro} brings the demo script and, finally, Section \ref{sec:conclusion} concludes this article and lists future works.

\section{NDN4IVC} \label{sec:proposal} 

NDN4IVC\footnote{https://github.com/insert-br/ndn4ivc} is a framework for simulation and experimentation of intelligent transportation systems and V-NDN applications. The environment uses two simulators well known in the literature, Ns-3, a network simulator based on discrete events, with ndnSIM module installed, and Sumo, a road-traffic and urban mobility simulator. The project integrates a comprehensive set of codes, models, and functionalities for more realistic simulation and helps the communication in vehicular information-centric networks. In addition to Vehicle-to-Vehicle communication (V2V), the installation and use of the Road Side Unit in the framework allows vehicles to communicate with other computer infrastructure (e.g., fog, edge, and cloud), known as V2I -- Vehicle-to-Infrastructure.

\begin{figure}[!hbt]
  \centering
  \includegraphics[width=.5\textwidth]{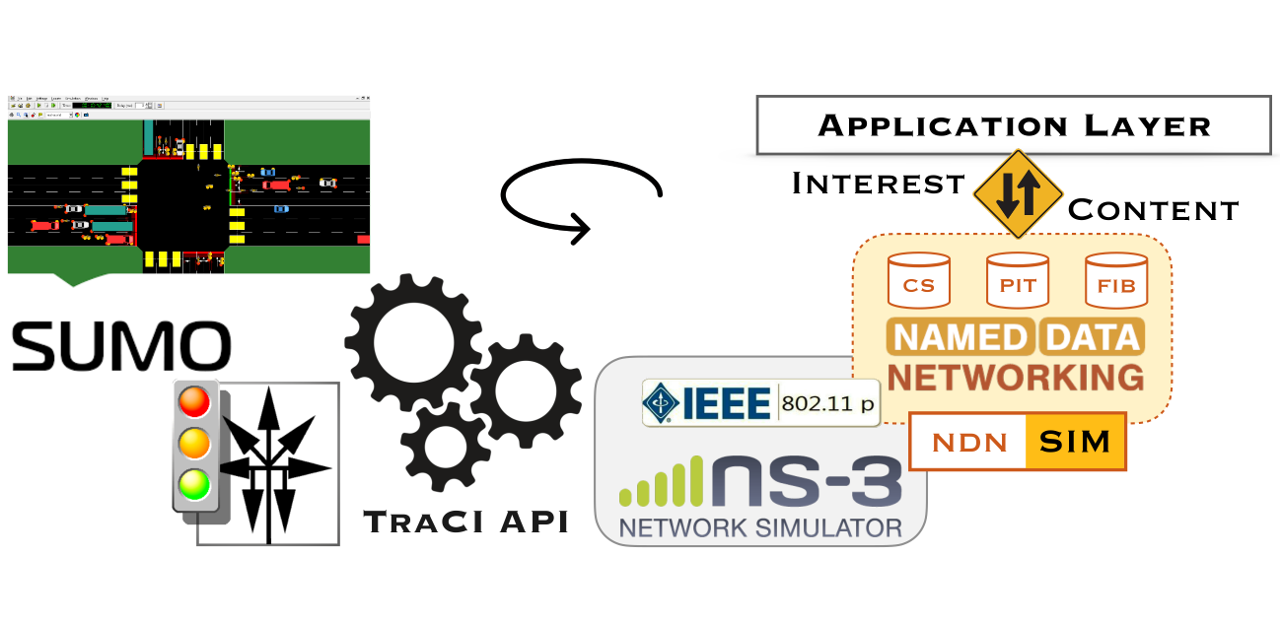}
  \caption{NDN4IVC - main components (overview)}
  \label{fig:componentes}
\end{figure}

Regarding the integration of the components in Figure \ref{fig:componentes}, it should be noted that Sumo and Ns-3 have different design characteristics, and each simulator interacts with the nodes in the environment differently. In Sumo, vehicles are created and removed during the entire simulation period, while Ns-3 does not support dynamic node insertion in the network. Therefore, the scenario must be previously described and statically created for each execution, specifying all nodes and parameters before starting the simulation. In summary, in the Ns-3, it is not possible to add and remove nodes during simulation natively. On the other hand, the simulation of vehicular networks requires the dynamic creation of nodes in the network because vehicles enter the scenario, follow a path (route) and leave the environment at the end of the trip. The module in \cite{VodafoneChair}, which also helps in this process, uses a node pool concept for Ns-3, with dynamic node startup and shutdown functions in the simulator. The idea is to activate and deactivate the device's resources on the network, based on the insertion or removal of the node by Sumo. In this case, a node in the network simulator with the network interface disabled, the application uninstalled and moved away from the range of other vehicles, will not receive, process, or create any new event in the environment that changes the simulation behavior. Thus, it is possible to make dynamic insertion and remove nodes during the simulation without making significant changes in the Ns-3 code. Vehicle insertion, removal, and mobility information are exchanged via TraCI interface \cite{10.1145/1400713.1400740}.

Figure \ref{fig:componentes} illustrates the main components of the framework. The bidirectional communication between Ns-3 and Sumo is the core, kernel, and cog of the system. This project uses the TraCI interface and a modified version of the module available in \cite{VodafoneChair} to integrate the entire environment.

\begin{figure*}[!hbt]
  \vspace{5pt}
  \centering
  \includegraphics[width=.99\linewidth, frame]{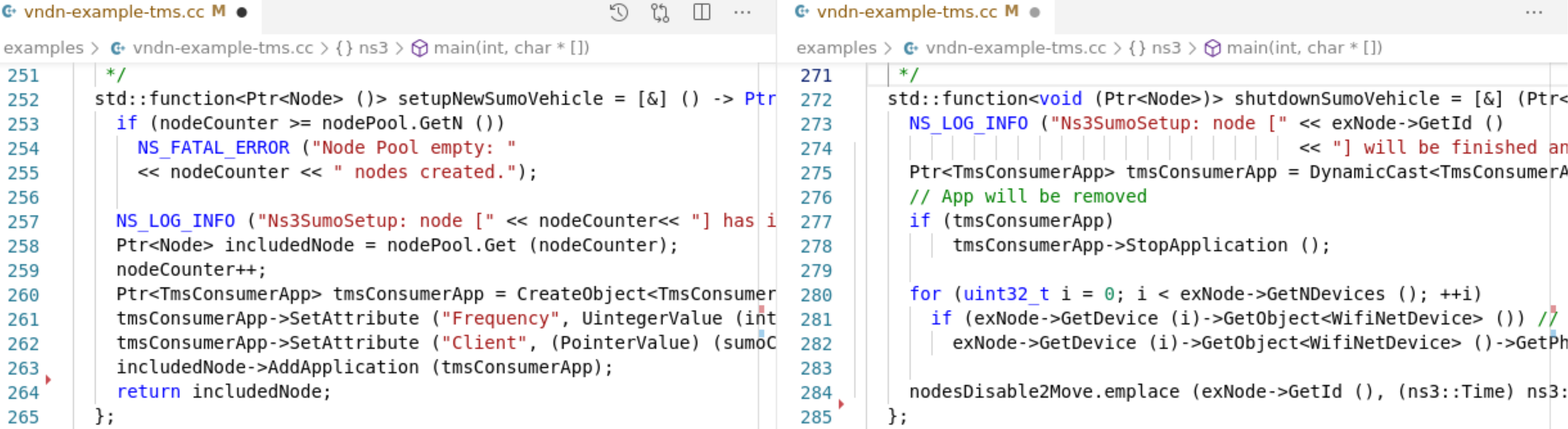}
  \caption{NDN4IVC - dynamic process for Ns-3 node startup and shutdown}
  \label{fig:code}
\end{figure*}

Figure \ref{fig:code} exhibits parts of the source code used by NDN4IVC to permit insertion of nodes during simulation. Lambda functions (\texttt{setupNewSumoVehicle} and \texttt{shutdownSumoVehicle}) are responsible for activating and deactivating nodes (dynamically) in the scenario synchronously with the mobility simulator (SUMO).

Table \ref{table:dir} presents the framework directory structure. The \texttt{doc} folder should be used for future and specific documentation. In the \texttt{examples} folder, simulation scenarios implemented in the environment are available. The \texttt{helper} and \texttt{model} directories contain  codes for proper simulation of vehicular named-data networking. Custom V-NDN forwarding strategies, sample applications, and help files for data link and physical layer configuration, via IEEE 802.11p, are available in these two directories.

In \texttt{Traces}, simulation scenarios for Sumo are stored. Users can modify the files to adapt the road traffic and mobility scenarios as needed. Therefore, it is important to note that new maps must follow the model present in the framework. Thus, it is possible to load new mobility scenarios into Ns-3 without adjusting the code. A \texttt{Makefile} is present in the sub-folders for each scenario to facilitate the synthetic mobility-generation process. Finally, the results folder must be used to store log files of experiments, data processing, and graphical presentation.

\begin{table}[!ht]
  \small
  \centering
  \caption{NDN4IVC directory structure}
  \label{table:dir}
  \begin{tabular}{|c|c|}
  \hline
  \textbf{\small Directory} & \textbf{\small Description}\\
  \hline
  \texttt{\small doc/} & {\small Additional documentation}\\
  \hline
  \texttt{\small examples/} & {\small Ns-3 simulation scenarios}\\
  \hline
  \texttt{\small helper/} & {\small Helper codes for configuration}\\
  \hline
  \texttt{\small model/} & {\small Applications and User Codes}\\
  \hline
  \texttt{\small traces/} & {\small Mobility data and road maps for Sumo}\\
  \hline
  \texttt{\small results/} & {\small Log files, data processing and graphics}\\
  \hline  
  \end{tabular}
\end{table}

\subsection{Use Cases}

Two use cases in V-NDN were defined to demonstrate possibilities for simulation and testing applications in NDN4IVC. For use case (i), only V2V communication is used and shows exchanging information between neighboring vehicles to optimize the flow of emergency vehicles in the left lanes. The use case (ii) implements a traffic management system (TMS) to exchange information about road congestion through V2X (Vehicle-to-everything) communication. In addition, this use case (ii) allows vehicles to change their routes to reduce travel time. All source code for these use cases can be viewed in the \texttt{model} and \texttt{example} directories in the project repository.

It is important to emphasize that these proposed scenarios have didactic objectives to demonstrate the framework and a customized environment for V-NDN simulation. Optimizations,  new use cases examples, and code contributions can be made through the official NDN4IVC repository on Github.

{\bf Use Case I}: traffic safety -- V2V communication. The application (i), road safety, periodically sends beacons of interest to propagate information about the vehicle's location to other neighboring nodes. Consequently, each node in the network is also able to maintain an up-to-date neighborhood table. Note that interest beacons are control messages, so there is no need to return any data packets. 

The proposed naming schema for the application (i) is  \texttt{\textbf{/}localhop\textbf{/}beacon\textbf{/}<node-id>\textbf{/}<road-id>\textbf{/}<pos- x>\textbf{/}<pos-y>\textbf{/}<pos-z>\textbf{/}<speed>}. Therefore, the scope for interest messages is localhop \cite{ndnlocalhop}. Thus, the NFD itself will control and limit the propagation of these messages in just one hop. Consequently, the application (i) does not have to avoid broadcast storms on the network since the NDN architecture provides this as a service for the application layer. The naming schema additionally specifies a message label, and when transmitting information about itself, each vehicle adds its id, vehicle type, location, and current speed to the name. Vehicles also process messages (beacons) received from other neighbors, and passenger vehicles should change to the right lane quickly, whether the message has been received from emergency vehicles. This use case uses a highway scenario, available in \texttt{traces/highway}.

{\bf Use Case II:} vehicular traffic efficiency -- V2X communication. The traffic management application, available in the framework, explores V2X communication. In this case, vehicles exchange messages and estimates about the level of congestion (average speed and occupancy rate) on the roads of interest. RSUs in the scenario are responsible for providing traffic information. Vehicles (consumers) send interest messages about the traffic conditions for all streets over the path. When congestion occurs upon any road above a threshold, vehicles can change the original route to find alternative pathways for a faster trip. In this use case (ii), vehicles also cooperate in the retransmission process on the network, playing the roles: data consumer, forwarder, when it is connected to either infrastructure or other vehicles, and data mule, when transporting data even without network connectivity, spreading content between different areas across the network. Unlike other mobile devices, vehicles have no concern with computational and storage capacity. A simple grid scenario was used for this application and is available in directory \texttt{traces/grid-map}.

The naming schema proposed for exchanging information in the application (ii) is \texttt{\textbf{/}service\textbf{/}traffic\textbf{/} <road-id>\textbf{/}<time-window>.} The naming schema discriminates the name of the service, the street (road) of interest, and the time window that represents the traffic conditions in the interval. To better data exchange in this scenario, the framework integrates and uses the JSON \texttt{Nlohmann C++}\footnote{https://github.com/nlohmann/json} library. Working with JSON format in the application layer is justified by being a compact, simple, intuitive, and easily described structure for exchanging data between systems and applications distributed on the network.

NDN4IVC makes adaptations in the multicast forwarding strategy available in the ndnSIM module to run applications more adequately for V-NDN context. By default, the multicast strategy sends back NACK messages when a given name is unreachable \cite{ndnnack}. In addition, due to design issues and simulation environment limitations, the ndnSIM module, when sending interest messages on the network, uses an event handling function that will process only one of the following conditions: (i) timeout, (ii ) receiving NACK or (iii) data. However, this technique is not suitable for V-NDN. It floods the network with NACK packets (No Route) due to the constant changes in the network topology and can make it impossible to receive data messages between neighboring nodes whether a NACK is processed first. The framework includes the multicast-vanet forwarding strategy to solve this issue. Other adjustments were also made to create NDN faces in ad-hoc mode, more appropriate for the IEEE 802.11p and vehicular named-data networking context.

\section{Evaluation} \label{sec:avaliacao}

This article evaluated the use case (ii). The experiments were planned through the complete factorial design \cite{books/daglib/0076234} to assess the work, with the following factors and respective levels: number of RSUs (1, 2), cache size (1, 1000), vehicle density (1000, 2000), and forwarding strategy (multicast, multicast-vanet). Vehicle density represents the number of vehicles per square kilometer, and the routes were randomly generated for each replication of the experiment. An accident event was generated, and the experiments use IEEE 802.11p standard with a communication range of 70 meters. The most relevant results will be discussed below, and the graphs have a confidence interval of 95\%.

\begin{figure}[!hbt]
  \centering
  \includegraphics[width=.5\textwidth]{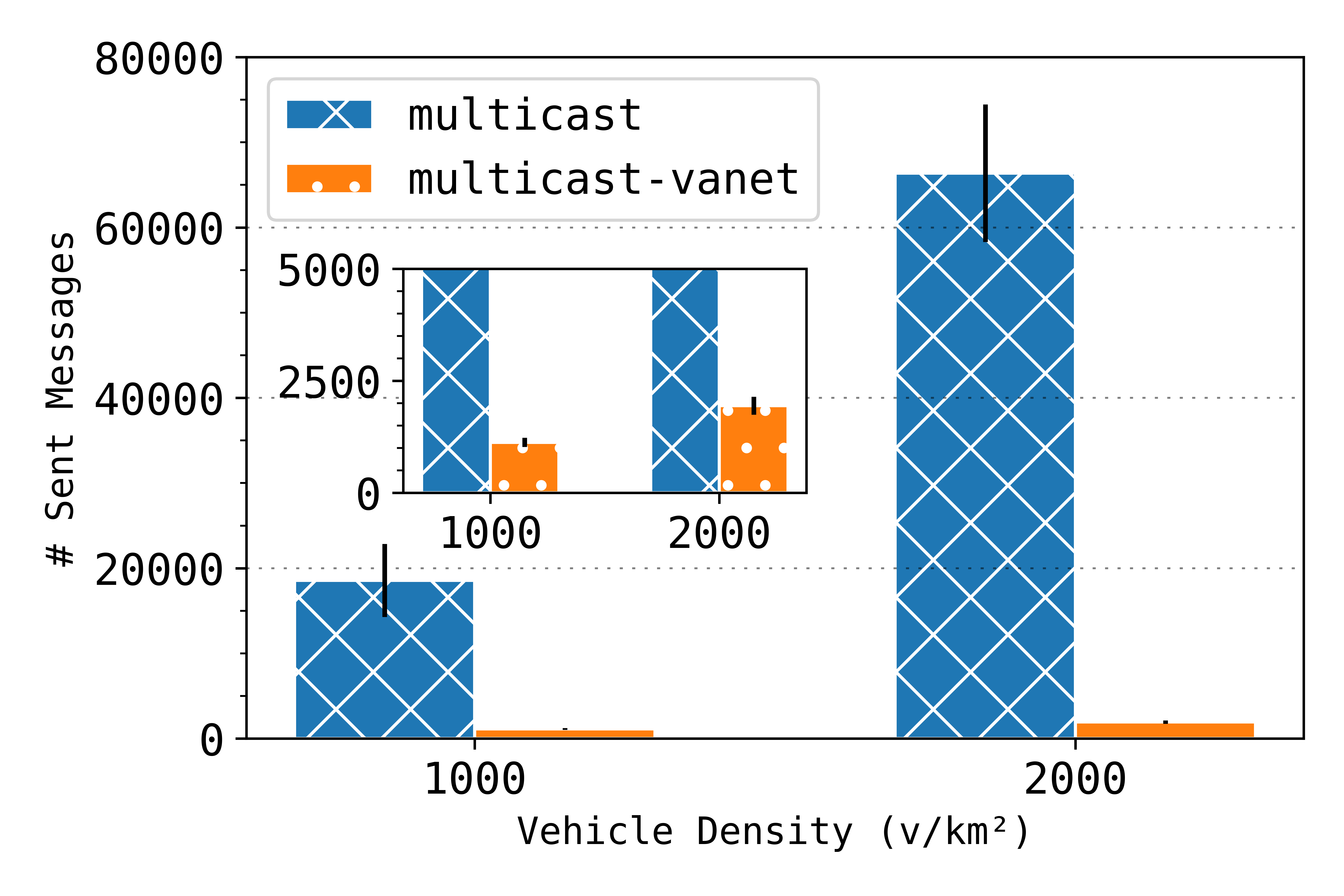}
  \caption{Number of packages sent by vehicle\\ density and forwarding strategy}
  \label{fig:res1}
\end{figure}

Figure  \ref{fig:res1} shows a significant increase in the number of packets sent on the network for the multicast forwarding strategy when the density of vehicles on the network increases. The excessive amount of NACK messages, informing that a given name is unreachable by a given node,  can explain this increase. 

For the multicast-vanet strategy, it is visible in Figure \ref{fig:res1} that the increase in the number of transmitted messages is low when doubling the number of vehicles. This behavior is due to the NDN stateful forwarding plane. The cache in the network layer and the Pending Interest Table (PIT) helps in a more optimized and consumer-oriented forwarding process, avoiding data redundancy and requests in the network.

\begin{figure}[!hbt]
  \centering
  \includegraphics[width=.5\textwidth]{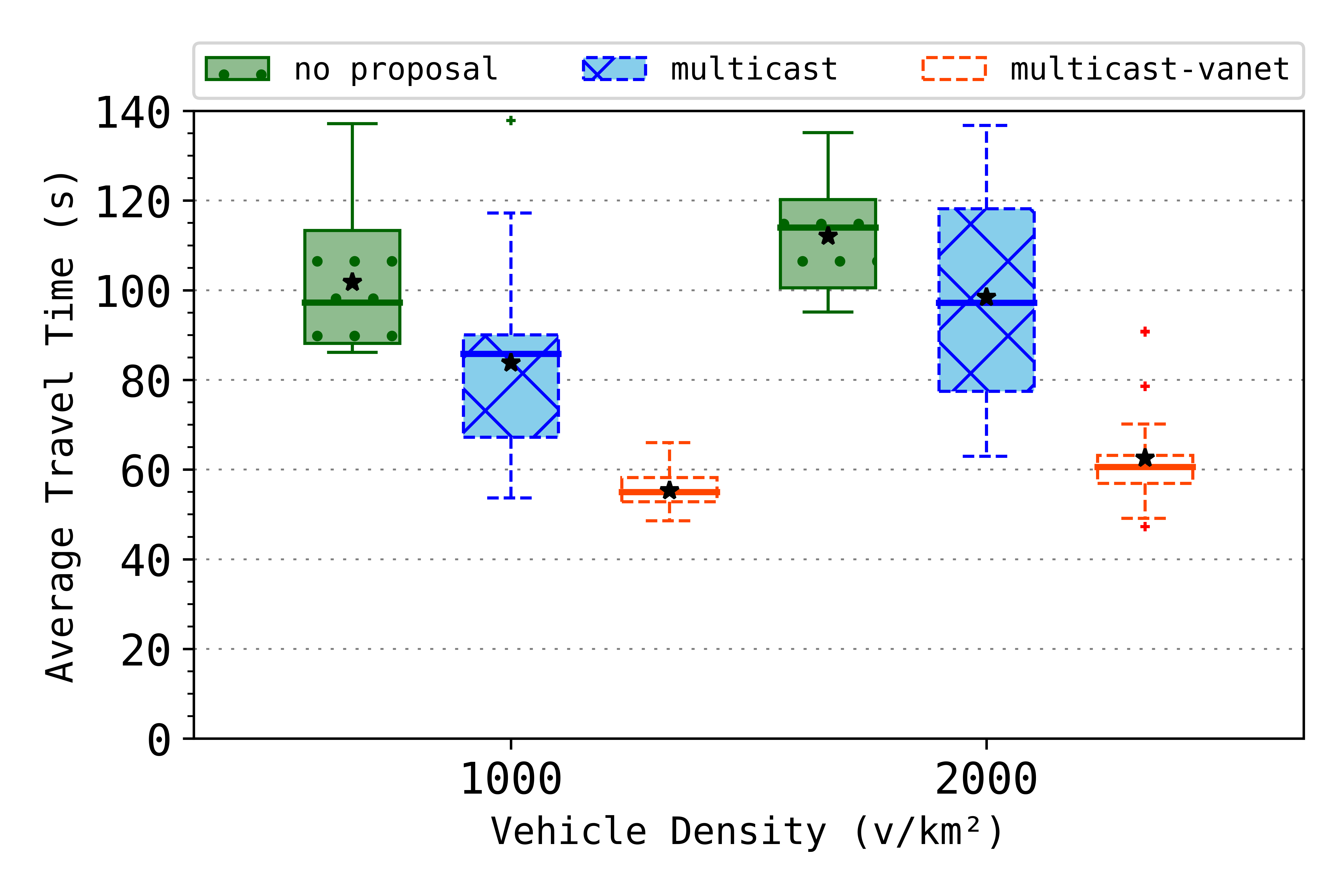}
  \caption{Average travel time by vehicle's\\ density and forwarding strategy}
  \label{fig:res2}
\end{figure}

Figure \ref{fig:res2} highlights the results for another metric evaluated, average travel time. In the scenarios without the proposal, the data were obtained directly by the Sumo simulator and are present for baseline and better evaluation of the results.  Notably, the vehicles' average travel time is longer in scenarios without the proposal. Despite improving the average travel time compared to scenarios without the proposal, the multicast forwarding strategy presents a considerable data spread, especially in experiments with higher vehicle density.

Scenarios with the multicast-vanet strategy have a better average travel time compared to the others. Furthermore, the results have low dispersion, indicating a significant improvement for all vehicles. This behavior remains even with an increase in the density of vehicles in the network.

Average travel time is a metric for the application and use case (ii) but shows that it is essential to use a customized and suitable environment for V-NDN simulating. For the same application, it is possible to obtain very different results, whether the environment is not previously validated and optimized for the context of the proposal.

\section{Demonstration} \label{sec:roteiro}

Sumo version 1.1.0 and Ns-3 version 3.30.1, with the ndnSIM module version 2.8, were used by NDN4IVC in version 1. The Ubuntu 18.04 LTS was used for testing and installing the entire environment, and the source code is available in the official repository on Github.

\begin{figure*}[!hbt]
    \vspace{20pt}
    \centering
    \subfigure[SUMO-GUI graphical interface]
    {
        \includegraphics[width=3.45in]{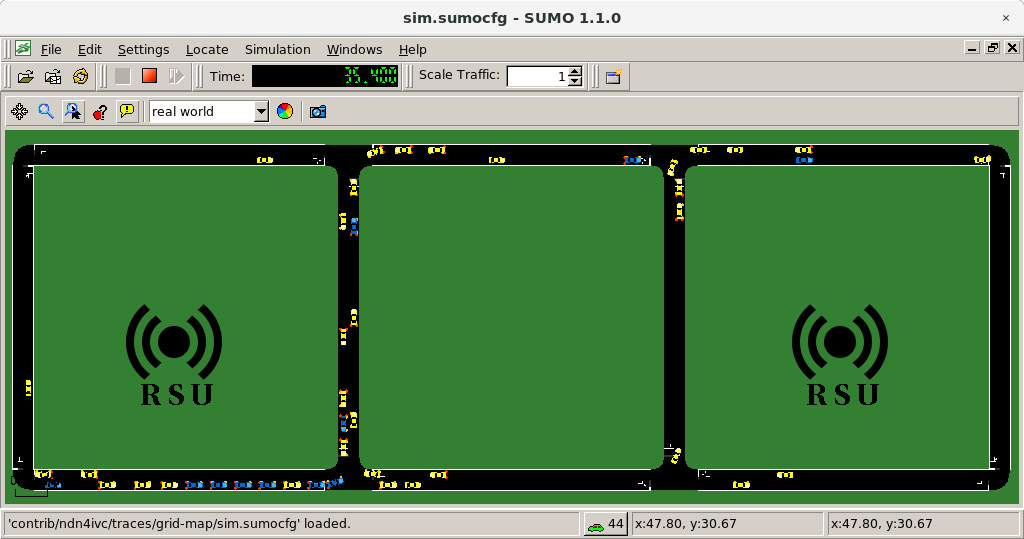}
        \label{fig:TPresult2}
    }
    \subfigure[PyViz interface in the Ns-3]
    {
        \includegraphics[width=3.45in]{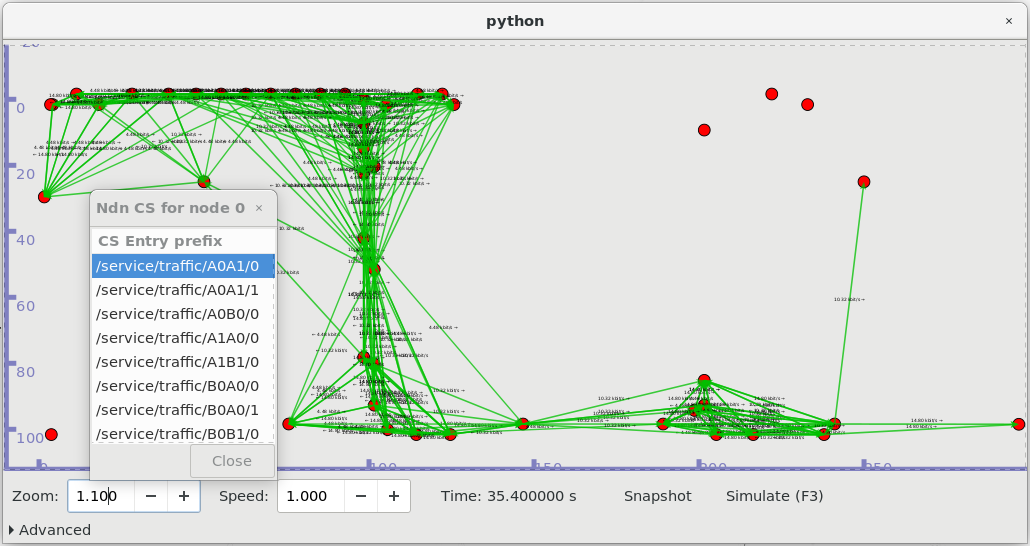}
        \label{fig:TPresult3}
    }
    \vspace{5pt}
    \caption{NDN4IVC -- graphical user interface (GUI) for live simulation}
    \label{fig:simrunning}
    \vspace{10pt}
\end{figure*}

Figure \ref{fig:simrunning} displays the application and use case (ii) running in the proposed environment. Figure \ref{fig:simrunning} (a) illustrates  Sumo's graphical interface with the mobility of vehicles on the road infrastructure. Finally, Figure \ref{fig:simrunning} (b) shows the communication between the respective nodes in the network. It is important to point out that the graphics module in Ns-3, developed in Python, by default displays the \texttt{y} axis inverted when compared to SUMO-GUI. However, it is just a visual issue, as the geographic coordinates in the plane are identical in both simulators. For better demonstration and user interaction, when vehicles process traffic information and select an alternative route, they change, in real-time, the default color to blue. It is more intuitive in the demo video\footnote{https://www.youtube.com/channel/UCzjOH9dSMyA5aoR-GZkAotw}.

\section{Conclusion} \label{sec:conclusion}

NDN4IVC customizes the entire environment and provides more efficient and adequate simulations of proposals in vehicular named-data networking (V-NDN) by integrating a network simulator and an urban mobility simulator. Two use cases were proposed to explore some possibilities, allowing to address and present the framework in practice. The results obtained from the simulation and the discussion presented address issues and problems presented in the article, demonstrating the framework's capacity and potential. 

As future works,  new scenarios and use cases that use others NDN architecture features, such as data security, will be added. Optimized forwarding strategies \cite{8379540} for vehicular named-data networks should also be implemented in future versions.

\section*{Acknowledgments} 
The authors would like to thank the partial support of CAPES (Brazilian Federal Agency for the Support and Evaluation of Graduate Education), FAPESB (Bahia State Research Support Foundation), and CNPq (National Council for Scientific and Technological Development).



\bibliographystyle{IEEEtran}
\bibliography{refs}

\end{document}